\documentclass[twocolumn]{revtex4}

\usepackage{graphicx}
\usepackage[usenames]{color}

\usepackage[english]{babel}

\usepackage[ansinew]{inputenc}
\usepackage{amsmath,amssymb,bm}
\usepackage{mathrsfs}
\usepackage[all,line]{xy}
\usepackage[T1]{fontenc}

\newcommand{\af}[1]{\!\left(#1\right)}
\newcommand{\tuborg}[1]{\left\{ #1 \right\}}
\newcommand{\kantpar}[1]{\left[ #1 \right]}

\renewcommand{\vec}[1]{\mathbf{#1}}                       

\newcommand{\mathlow}[1]{_{\mathrm{#1}}}

\newcommand{\Exp}[1]{e^{#1}}

\newcommand{\pardd}[3][]{\frac{\partial^{#1} #2}{\partial #3^{#1}}}
\newcommand{\dd}[3][]{\frac{d^{#1} #2}{d #3^{#1}}}

\newcommand{\parddt}[3][]{\partial^{#1} #2 / \partial #3^{#1}}

\newcommand{\Max}{\mathlow{max}}

\newcommand{\Tot}{\mathlow{tot}}
\newcommand{\Res}{\mathlow{res}}

\newcommand{\defi}{\equiv}

\newcommand{\kL}{k_L}

\newcommand{\Atom}{a}
\newcommand{\Molecule}{m}
\newcommand{\Type}{\sigma}
\newcommand{\SubAtom}{_\Atom}
\newcommand{\SubMolecule}{_\Molecule}
\newcommand{\SubType}{_\Type}

\newcommand{\SpinUp}{\uparrow}
\newcommand{\SpinDown}{\downarrow}

\newcommand{\NumSites}{M}

\newcommand{\Num}{N}
\newcommand{\NumTot}{N\Tot}

\newcommand{\NumA}{N\SubAtom}

\newcommand{\NumM}{N\SubMolecule}

\newcommand{\Energi}{E}

\newcommand{\EM}{\Energi\SubMolecule}
\newcommand{\E}{\Energi}

\newcommand{\EAnum}[1]{\Energi\SubAtom^{#1}}
\newcommand{\EMnum}[1]{\Energi\SubMolecule^{#1}}
\newcommand{\Einum}[1]{\Energi\SubType^{#1}}

\newcommand{\KemPot}{\mu}

\newcommand{\FF}{\eta}
\newcommand{\FFC}{\FF_c}

\newcommand{\Eres}{\E\Res}

\newcommand{\EF}{E\mathlow{F}}

\newcommand{\ERi}{E_{R,\Type}}
\newcommand{\ERA}{E_{R,\Atom}}

\newcommand{\TF}{T\mathlow{F}}

\newcommand{\f}{f}
\newcommand{\fA}{\f\SubAtom}
\newcommand{\fM}{\f\SubMolecule}

\newcommand{\fAnum}[1]{\f\SubAtom^{#1}}
\newcommand{\fMnum}[1]{\f\SubMolecule^{#1}}
\newcommand{\fAr}[1]{\f\SubAtom^{#1}}
\newcommand{\fMr}[1]{\f\SubMolecule^{#1}}

\newcommand{\dens}{\rho}
\newcommand{\dA}{\dens\SubAtom}
\newcommand{\dM}{\dens\SubMolecule}

\newcommand{\kB}{k\mathlow{B}}

\newcommand{\refEq}[1]{(\ref{#1})}

\newcommand{\Pot}{V}

\newcommand{\EresMaxS}{E^{\Entropy\Max}\mathlow{res}}

\newcommand{\EOneD}{\mathcal{E}}

\newcommand{\GapThreeD}[2][\Type]{\Delta_{#1}}
\newcommand{\BandWidthThreeD}[2][\Type]{\delta_{#1}}

\newcommand{\Ti}{T_i}
\newcommand{\Tf}{T_f}
\newcommand{\Sf}{\Entropy_f}
\newcommand{\Si}{\Entropy_i}

\newcommand{\Entropy}{S}
\newcommand{\STot}{\Entropy\Tot}
\newcommand{\SA}{\Entropy\SubAtom}
\newcommand{\SM}{\Entropy\SubMolecule}

\newcommand{\ddconst}[4][]{\af{\dd[#1]{#2}{#3}}_{#4}}

\newcommand{\SAP}{\SA^p}
\newcommand{\SMP}{\SM^p}

\newcommand{\LatticePot}{L}

\newcommand{\abs}[1]{\lvert #1 \rvert}

\renewcommand{\figurename}{Figure}

\def\ket#1{\mathinner{|{#1}\rangle}}

\let\protect\relax
{\catcode`\|=\active
\xdef\Braket{\protect\expandafter\noexpand\csname Braket \endcsname}
\expandafter\gdef\csname Braket \endcsname#1{\begingroup
   \ifx\SavedDoubleVert\relax
     \let\SavedDoubleVert\|\let\|\BraDoubleVert
   \fi
   \mathcode`\|32768\let|\BraVert
   \left\langle{#1}\right\rangle\endgroup}
}
\def\BraVert{\@ifnextchar|{\|\@gobble}
   {\egroup\,\mid@vertical\,\bgroup}}
\def\BraDoubleVert{\egroup\,\mid@dblvertical\,\bgroup}
\let\SavedDoubleVert\relax

{\catcode`\|=\active
\xdef\set{\protect\expandafter\noexpand\csname set \endcsname}
\expandafter\gdef\csname set \endcsname#1{\mathinner
      {\lbrace\,{\mathcode`\|32768\let|\midvert #1}\,\rbrace}}
\xdef\Set{\protect\expandafter\noexpand\csname Set \endcsname}
\expandafter\gdef\csname Set \endcsname#1{\left\{%
   \ifx\SavedDoubleVert\relax \let\SavedDoubleVert\|\fi
   \:{\let\|\SetDoubleVert
   \mathcode`\|32768\let|\SetVert
   #1}\:\right\}}
}
\def\midvert{\egroup\mid\bgroup}
\def\SetVert{\@ifnextchar|{\|\@gobble}
  {\egroup\;\mid@vertical\;\bgroup}}
\def\SetDoubleVert{\egroup\;\mid@dblvertical\;\bgroup}

\begingroup
\edef\@tempa{\meaning\middle}
\edef\@tempb{\string\middle}
\expandafter \endgroup \ifx\@tempa\@tempb
\def\mid@vertical{\middle|}
\def\mid@dblvertical{\middle\SavedDoubleVert}
\else
\def\mid@vertical{\mskip1mu\vrule\mskip1mu}
\def\mid@dblvertical{\mskip1mu\vrule\mskip2.5mu\vrule\mskip1mu}
\fi

\begin{document}
\title{Adiabatic cooling of a tunable Bose-Fermi mixture in an optical lattice}
\author{O. S{\o}e S{\o}rensen}
\affiliation{Lundbeck Foundation Theoretical Center for Quantum System Research, Department of Physics and Astronomy, University of Aarhus, DK-8000 {\AA}rhus C, Denmark}
\author{P. B. Blakie}
\affiliation{Jack Dodd Centre for Photonics and Ultra-Cold Atoms, Department of Physics, University of Otago, New Zealand}
\author{N. Nygaard}
\affiliation{Lundbeck Foundation Theoretical Center for Quantum System Research, Department of Physics and Astronomy, University of Aarhus, DK-8000 {\AA}rhus C, Denmark}

\begin{abstract}
We consider an atomic Fermi gas confined in a uniform optical lattice potential, where the atoms can pair into molecules via a magnetic field controlled narrow Feshbach resonance.  
Thus by adjusting the magnetic field the portion of fermionic and bosonic particles in the system can be continuously varied. We analyze the statistical mechanics of this system and consider the interplay of the lattice physics with the atom-molecule conversion.
We study the entropic behavior of the system and characterize the temperature changes that occur during adiabatic ramps across the Feshbach resonance. We show that an appropriate choice of filling fraction can be used to reduce the system temperature during such ramps. 

\end{abstract}
\maketitle

\section{Introduction}
\begin{figure}[b]
   \centering
   \includegraphics[width=\columnwidth]{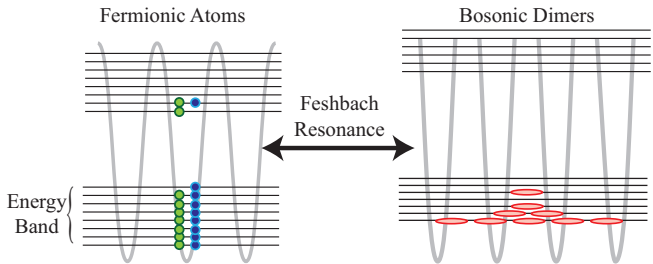}
   \caption{Schematic diagram of the system under consideration: Fermionic atoms in two different spin states occupy states in an optical lattice and are coupled by a  Feshbach resonance into bosonic dimers. Because the bosonic dimers have a larger polarizability they experience a deeper lattice potential with a different spectrum to that for the atoms.}
   \label{fig:diagram}
\end{figure}

One of the most fascinating consequences of many-body quantum theory is the striking difference in behavior between systems made from identical bosons or fermions at low temperatures. 
The use of Feshbach resonances in atomic systems is now a routine technique for pairing atoms into diatomic molecules~\cite{Donley2002, Regal2003, Herbig2003, Jochim2003, Cubizolles2003, Zwierlein2003, Strecker2003, Durr2004}, and leads to the intriguing scenario:  the controlled conversion of an ultra-cold degenerate Fermi gas into a system of bosonic dimers. 
Hence, in experiments it is possible to control the statistics of the system by tuning across the Feshbach resonance. In this paper we consider a particular case of such a system confined in an optical lattice, schematically shown in \figurename~\ref{fig:diagram}. Atom-molecule conversion in an optical lattice has been demonstrated in several recent experiments~\cite{Moritz2005, Stoferle2006, Kohl2006, Chin2006, Thalhammer2006, Ospelkaus2006}. We develop formalism to quantify the effect that ramping across the Feshbach resonance has on the temperature of the system in the adiabatic limit by developing expressions for the entropy in the nearly degenerate regime.

The essence of a Feshbach resonance is that free atom pairs are coupled to a discrete dimer state in a closed scattering channel. The energy of this closed channel bound state is detuned from the atomic threshold by an amount $\Eres$, which is tunable with an applied magnetic field, since the unbound atoms and the closed channel dimer have different magnetic moments (e.g. see \cite{Kohler2006}). If $\Eres<0$ the closed channel bound state corresponds to a bound molecular state of the coupled system.
When $\Eres>0$ the discrete state is embedded in the atomic continuum and has a finite lifetime. Consequently, it manifests itself not as a true bound state, but as a scattering resonance. However, if the width of the Feshbach resonance is sufficiently narrow, the lifetime of the closed channel dimer state is long enough that we may consider it a quasi-bound state of the system. Hence, characterizing both the bound and the quasi-bound state of an atom pair as molecules, the number of unbound atoms and molecules in the gas are well-defined at any given instant.  

The thermodynamics of such a Feshbach-resonant Fermi gas have been studied in free space~\cite{Carr2004} as well as in a harmonic confinement~\cite{Williams2004,Watabe2007}, and in previous work we have considered the case of a lattice potential and analyzed the behaviour of the chemical potential and the atomic and molecular  populations in the degenerate regime ($\kB T \lesssim 0.1 \EF$). There has also been a {great} deal of work on systems with broad Feshbach resonances, which require a full many-body theory for resonantly interacting Fermi atoms~\cite{Haussmann2007, Haussmann2008, Hu2006a, Hu2006b, He2007, Perali2004}. For a deep lattice this has been studied at zero temperature by Koetsier et al.~\cite{Koetsier2006}.

We believe that the optical lattice introduces considerable new physics~\cite{Moritz2005, Winkler2006, Syassen2007, Nygaard2008a, Nygaard2008b}, and our ambition here is to understand the resulting changes in the thermodynamics by considering a simple model for the system:  the case of ideal particles (i.e. we neglect interactions for both the fermionic atoms and bosonic dimers)  in a translationally invariant lattice potential. 
In this model the only effect of the Feshbach resonance is to maintain chemical equilibrium between the two species, while a major simplification, this captures the essential physics of atom-molecule conversions in experiments~\cite{Williams2006, Chin2004}, and can be shown to be exact in the limit where the resonance is infinitey narrow~\cite{Gurarie2007}.  We note that for a (finite) narrow Feshbach resonance the inclusion of interactions should only impact our results quantitatively.

In this paper we look at the entropic behavior of the system with special focus on the low-temperature regime, since isentropic surfaces  characterize what will happen to a system in an adiabatic process. Such adiabatic processes are often used in experiment, and our particular case of interest here is to predict how slowly ramping   $ \Eres$ across the Feshbach resonance will effect the temperature, and hence the degeneracy of the system. By looking at the temperature variation along such curves we can identify in which regimes cooling will occur, and surprisingly this turns out to depend on the average number of atoms per site in the optical lattice.
Lattice physics plays an important role in changing the thermal properties of the system, indeed  similar studies to what we present here have been undertaken for pure ideal Bose and Fermi gases in the translationally invariant lattice \cite{Blakie2004a,Blakie2005a} and more recently for the case with  additional  harmonic confinement \cite{Kohl2006a,Blakie2007a,Blakie2007b}. 
Interactions play an important role in deep lattices, and more recent work has examined the effect these interactions have on the thermal excitation generated during the preparation of bosonic Mott-insulating states \cite{Ho2007a,Gerbier2007a}, and on the feasibility of achieving the fermionic Nèel state \cite{Koetsier2008a}. 
Finite temperature mixtures of atomic Bose and Fermi gases (with no conversion mechanism) in lattices have been studied \cite{Cramer2008a} in an attempt to explain recent experiments \cite{Ospelkaus2006a}.

\section{Formalism}\label{sec:formalism}
We consider a dilute system of Fermi atoms of mass $m\SubAtom$ in a  simple cubic optical lattice potential with $\NumSites^3$ sites and subject to periodic boundary conditions. The total number of atoms, $\NumTot$, are divided equally into two different internal states which we denote as $\ket \SpinUp$ and $\ket \SpinDown$, with populations $\Num^\SpinUp$ and $\Num^\SpinDown$, respectively. We define the filling fraction $\FF$ to be the average number of each type of atom on each site
\begin{align}
\label{eq:FFdef}
\FF \defi \frac{\Num^\SpinUp}{\NumSites^3}
= \frac{\Num^\SpinDown}{\NumSites^3}
= \frac{\NumTot}{2\NumSites^3} .  
\end{align}
By applying a tunable magnetic field we  can make the two atomic species interact with a Feshbach resonance and control whether these particle exist as individual (unbound) atoms or participate in bosonic dimers with mass $m\SubMolecule = 2 m\SubAtom$. We assume that the atoms and diatomic molecules are in thermal and chemical equilibrium, and we define the mean number of atoms ($\NumA$) and molecular dimers ($\NumM$) with the number conservation condition
\begin{align}
\NumTot = \NumA + 2 \NumM .
\label{eq:NumCons}
\end{align}

\subsection{Thermodynamics}
The thermal equilibrium condition for our system ensures that the atoms and the molecules share the temperature $T$, and that the chemical potentials of the atoms and molecules are related as $\KemPot\SubMolecule = \KemPot\SubAtom^\SpinUp + \KemPot\SubAtom^\SpinDown$. Since we are considering an equal spin-mixture, this simplifies to
\begin{align}
\KemPot\SubMolecule = 2 \KemPot ,
\end{align}
where $\KemPot \defi \KemPot\SubAtom^\SpinUp = \KemPot\SubAtom^\SpinDown$ is the common atomic chemical potential, and $ \KemPot\SubMolecule$ is the molecular chemical potential.
As the atoms only interact through the Feshbach resonance, we can describe the system by the single-particle energy levels, and the occupation of these are given by the Fermi-Dirac and Bose-Einstein distributions for the atoms and the molecules, respectively. Taking into account that we can effectively shift the molecular energy spectrum by an amount $\Eres$ relative to the atomic spectrum via the Feshbach resonance, the relevant distributions are
\begin{subequations}
\label{eq:FDBE}
\begin{align}
 \fAr{r} &\defi \frac{1}{\Exp{(\EAnum{r} - \KemPot)/\kB T} + 1} , \\
 \fMr{r} &\defi \frac{1}{\Exp{(\EMnum{r} + \Eres - 2\KemPot)/\kB T} - 1} .
\end{align}
\end{subequations}
The quantities $\EAnum{r}$ and $\EMnum{r}$ denote the single particle energy levels in the lattice for the atoms and molecules respectively, where $r$ is an appropriate quantum number ($r = 0,1,2,\ldots$). Since the molecules are governed by Bose-Einstein statistics we have the usual constraint that the (molecular) chemical potential must lie below the lowest molecular single particle state, i.e.  $\KemPot\SubMolecule < \EMnum{0} + \Eres$. For each choice of the resonance energy, filling fraction, temperature and lattice depth the chemical potential $\KemPot$ is then determined from the conservation of the total particle number (\ref{eq:NumCons}) where
\begin{subequations}
\label{eq:NumPart}
\begin{align}
 \NumA &= 2 \sum_{r = 0}^\infty \fAr{r}, \\
 \NumM &= \sum_{r = 0}^\infty\fMr{r} .
\end{align}
\end{subequations}
Once the relation between $\KemPot$, $T$ and the externally adjustable parameters has been established, it is straight-forward to calculate any thermodynamic quantity such as the entropy~\cite{LandauLifshitz}  
\begin{subequations}
\label{eq:entropy}
\begin{align}
 \STot &=\SA+\SM,\\
 \SA &= - 2  \sum_{r=0}^{\infty} \kantpar{\fAr{r} \ln{\fAr{r}} + (1 - \fAr{r}) \ln(1 - \fAr{r})} \kB ,  \\
 \SM &= - \sum_{r=0}^{\infty} \kantpar{\fMr{r} \ln{\fMr{r}} - (1 + \fMr{r}) \ln(1 + \fMr{r})} \kB,
\end{align}
\end{subequations}
where $\STot$ is the total system entropy, which is additively formed from the atomic ($\SA$) and molecular ($\SM$) subsystem entropies.

\subsection{Energy levels in the lattice}\label{sec:E_in_lattice}
The lasers which create the optical lattice have the wavelength $\lambda_L = 2\pi/\kL$ and the resulting potential is $\Pot\SubType\af{\vec x} = \Pot_{0,\Type} \LatticePot(\vec x)$ with $\Type= \Atom,\Molecule$ for atoms and molecules, respectively, and where
\begin{align*}
\LatticePot(\vec x)
= \sin^2\af{\kL x} + \sin^2\af{\kL y} + \sin^2\af{\kL z},
\end{align*}
is the dimensionless shape of the lattice. The potential depth $\Pot_{0,\Type}$ for the two different species is not the same, since a molecule consists of two atoms and therefore experiences twice the Stark shift of an individual atom: $\Pot_{0,m} = 2\Pot_{0,a}\defi 2\Pot_0$.

To find the possible energy levels we solve the stationary Schr\"odinger equation, which for our simple cubic lattice separates into three 1D Bloch equations. To find the 3D energies it is therefore sufficient to calculate the 1D energy levels and make all possible triplets thereof. It is convenient to rescale the stationary Schr\"odinger equation for the species $\Type$ by the corresponding recoil energy $\ERi \defi \hslash^2 \kL^2 / 2m\SubType$:
\begin{align}
\tuborg{- \frac{\hslash^2}{\kL^2} \nabla^2 + \bar\Pot_{0,\Type}\LatticePot(\vec x)} \psi^r_\Type(\vec x)  &=  \bar E^r_\Type {r} \psi^r_\Type(\vec x),
\end{align}
where the barred quantities are energies in units of $\ERi$. The operator on the left hand side only depends on the $\Type$ through the species-dependent lattice potential $\Pot_{0,\Type}$ and therefore the spectrum can be written as $\bar E_\Type^r = \bar E^r(\bar \Pot_{0,\Type})$ for some species-independent function $\bar E^r$. Furthermore, since the molecules are twice as heavy as the atoms and feel twice as deep a potential, we have the relation $\bar \Pot_{0,\Molecule} = 4 \bar \Pot_{0,\Atom}$, which means that the energy spectra for $\bar\Pot_{0,\Molecule}$ and $4 \bar \Pot_{0,\Atom}$ have precisely the same shape but with different scalings on the energy axis.

\begin{figure}[htpb]
    \centering
    \includegraphics[width=\columnwidth]{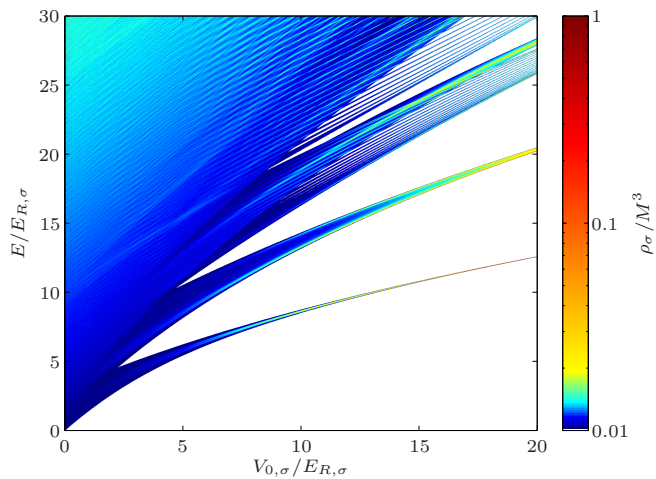}
    \caption{The density of states for a $31 \times 31 \times 31$ optical lattice as a function of the potential depth for the particle type $\Type$. The density of states is found by binning the energy levels in intervals of width $0.04 \ERi$.}
    \label{fig:DOS3D2}
\end{figure}

\subsubsection{Energy scales}
The fact that the atoms and molecules experience different lattice potentials thus leads to a relative shift of the two spectra in addition to the magnetic field adjustable Feshbach detuning $\Eres$. Consequently, the position and width of the Feshbach resonance, as indicated by the interconversion of atoms and molecules, depends on the depth of the optical lattice potential \cite{Ole2008a}. 

In \figurename~\ref{fig:DOS3D2} we illustrate the energy spectrum in the 3D lattice by plotting the density of states as a function of the lattice depth. The energy levels are distributed in a band-structure, where the bandwidths and the band gaps are strongly dependent on the depth of the lattice. The width of the allowed energy bands decreases as the lattice becomes deeper and for very deep lattice potentials the lowest energy band approximately reduces to a single energy level with a degeneracy that scales with the size of the system. On the other hand, the gaps between energy bands increase as the potential depth increases.

Because of the different potentials the atoms and molecule experience, there is an energy splitting between the lowest atomic and molecular levels in the optical lattice. Using a tight binding analysis, appropriate to the regime $V_0\gtrsim4 \ERA$, we find that this splitting is given by
\begin{align}
\label{eq:EnergySplit}
\EMnum{0} - \EAnum{0} \approx \frac{3}{8}\af{1 + \frac{3}{8}\frac{1}{\sqrt{\Pot_{0,\Atom} / \ERA}}} \ERA ,
\end{align}
as derived in Ref. \cite{Ole2008a}.

In a 1D periodic potential a band gap opens at the edges of the first Brillouin zone for an arbitrarily small lattice depth. On the contrary, in a 3D system the continuum is only broken up into separated bands if the lattice potential is sufficiently deep, as is clear from \figurename~\ref{fig:DOS3D2}.
In general it is of interest to know the width of the  ground  energy band  ($\BandWidthThreeD{0}$) and  the gap between the ground and first excited energy bands ($\GapThreeD{0}$) in the lattice.
We find that
\begin{align}
\GapThreeD{0}  &\approx \af{2\sqrt{\frac{\Pot_{0,\Type}}{\ERi}} - 1 - \frac{23}{24} \frac{1}{\sqrt{\Pot_{0,\Type} / \ERi}}} \ERi ,\\
\BandWidthThreeD{0}& \approx\frac{48}{\sqrt{\pi}}\af{\frac{\Pot_{0,\Type}}{\ERi}}^{3/4}\exp\af{-2\sqrt{\frac{\Pot_{0,\Type}}{\ERi}}}\ERi,
\end{align}
where both results are valid in the tight binding regime with the first result derived in Appendix \ref{AppndE}, and the second result is obtained from the harmonic oscillator approximation for the tunneling matrix element.
We remark that $\GapThreeD{0}/\kB$ sets the temperature scale for the excited bands to be relevant for the thermodynamics of particle to type $\Type$ in the system. 

A convenient energy scale is the Fermi energy $\EF\af{\FF, \Pot_0}$, which is taken to be the highest occupied energy level, when all the atoms are unbound. We note that for our choice of energy origin the relevant Fermi temperature for characterizing degeneracy is given by $\TF = (\EF - \EAnum{0} )/ \kB$, with $\EAnum{0}$ the atomic ground state energy. If $\FF < 1$ ther Fermi energy lies in the ground band (for sufficiently low temperatures) and the degeneracy condition is $\kB T \lesssim \BandWidthThreeD[\Atom]{0}$.

\begin{figure}[tpb]
 \centering
 \includegraphics[width=\columnwidth]{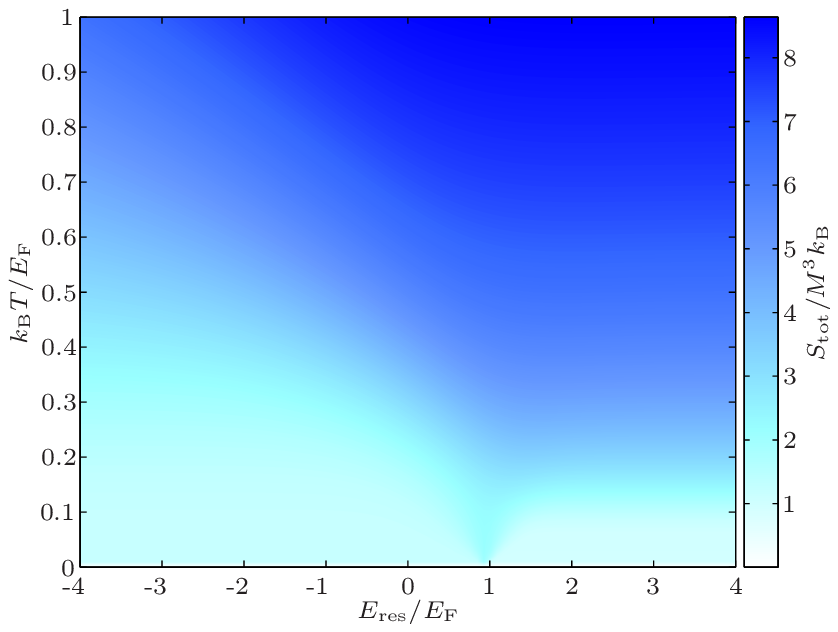}
 \caption{The system entropy $\STot$  in the $(\Eres , T)$ plane for $\Pot_0 = 10\ERA$ and $\FF = 0.8$.}
\label{fig:Entropygraphs}
\end{figure}

\section{Entropy}
\label{sec:Entropy}

To characterize the state properties of our system we numerically calculate the chemical potential for various points in the phase space $(\Pot_0 , \FF , \Eres , T)$, as discussed in Sec. \ref{sec:formalism}. Of most interest to us here is the behavior of the system entropy (\ref{eq:entropy}) as $\Eres$ and $T$ vary, for which a typical example is shown in \figurename~\ref{fig:Entropygraphs}. We make the following general observations on this dependence:
\begin{description}

\item[Atomic regime:] Well above the resonance, i.e. for $\Eres \gtrsim 2 \EF$, the entropy is independent of $\Eres$. In this regime the lowest molecular energy levels are at a much higher energy than the atomic levels and are inaccessible. This result is only valid in the regime where $\kB T<\EF$, so that we can neglect thermal activation of the molecular states.
\item[Molecular regime:] Well below the resonance, i.e. for $\Eres \lesssim -2\EF$, the opposite regime is entered into, in which  only the molecular states are accessible at low temperatures. Hence, in the lower left hand corner of the phase diagram the entropy is almost independent of $\Eres$.  At higher temperatures the resonance energy plays a larger role, since the molecules start dissociating at a temperature on the order of $\abs{\Eres} / \kB$.
\item[Entropy peak:] A notable  feature of the entropy phase diagram is the behavior at $\Eres \approx \EF$ where the entropy as a function of the resonance energy has a maximum along each isotherm -- a feature most noticeable  at low temperatures. We investigate this feature further below, but note that it arises where the system is an equal mixture of atoms and molecules, and thus able to obtain maximum disorder for a given temperature.
\end{description}

\begin{figure*}[htpb]
    \centering
    \includegraphics[width=0.8\textwidth]{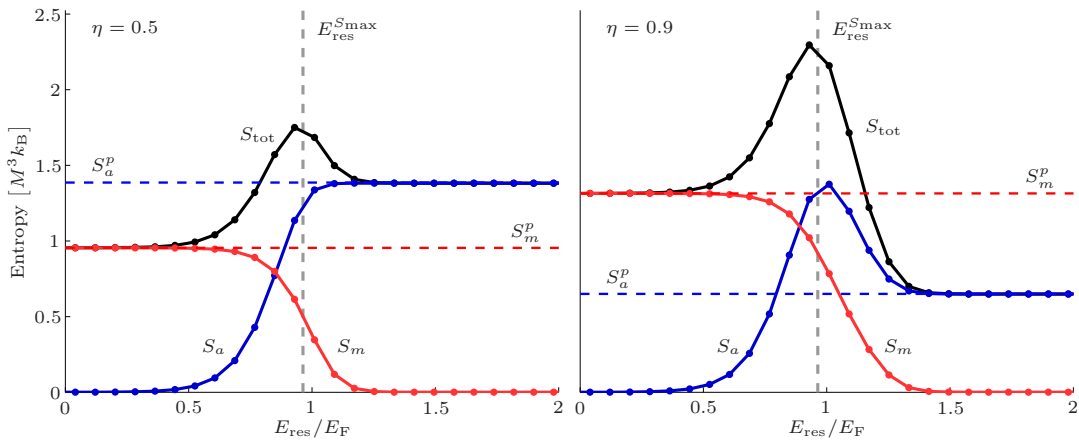}
    \caption{The atomic and molecular contributions to the entropy as well as the total entropy as functions of $\Eres$ at constant temperature, $\kB T = 0.04\EF$, lattice depth $\Pot_0 = 10 \ERA$ and for two different filling fractions. The horizontal dashed lines are the atomic and molecular entropy plateaus, Eqs. (\ref{eq:LimitEntropyA}) and (\ref{eq:LimitEntropyM}), respectively, and the vertical dashed lines indicate the resonance energy maximizing the entropy as approximated by (\ref{eq:EresSmaxApprox}). }%
    \label{fig:EntropyTop}
\end{figure*}

\subsection{Entropy peak}\label{sec:EntroPeak}
To analyze this last effect in more detail we look at the contributions to the entropy from both the atoms and the molecules along lines of constant temperature for $k_{\rm{B}}T\lesssim 0.1 \EF$. We will show that when approaching the transition zone, where the conversion between atoms and molecules takes place, the entropy increases from either side.

To clarify the underlying physics we develop a simple analytic model that should provide a good description in the deep lattice limit where the lowest energy bands are flat, provided the temperature is small compared to the first band gap but large compared to the ground bandwidth, i.e.  
\begin{align}
\label{eq:Econdition}
\BandWidthThreeD{0} \ll \kB T \ll \GapThreeD{0}.
\end{align}
We also require that $\FF \leq 1$ so that higher bands are not populated by atoms due to the Pauli exclusion principle, and furthermore that a condensate does not occur in the molecular system, which would require a unique ground state. 
We note that these conditions are broadly consistent with the typical regime that experiments operate in.
In this case we can approximate the density of states for the atoms and molecules as
\begin{subequations}
\begin{align}
  \dA(E) &= 2\NumSites^3\delta(E-\EAnum{0}),\\
  \dM(E) &= \NumSites^3\delta(E-\EMnum{0}-\Eres),
\end{align}
\end{subequations}
where, in the validity regime of this model, we can use  (\ref{eq:EnergySplit})  to relate the difference between $\EAnum{0}$ and $\EMnum{0}$ to the lattice depth.

In this approximation the level occupations (of the ground band) are independent of the particular level under consideration, and depend only on whether the particle is an atom or a molecule, for which we denote the mean level occupation as $\fAr{0}$ and $\fMr{0}$, respectively. The corresponding entropy contributions will be
\begin{subequations}
\label{eq:SingleLevelEntropy}
\begin{align}
 \SA &= - 2 \NumSites^3 \kantpar{\fAr{0} \ln \fAr{0} + (1 - \fAr{0}) \ln(1 - \fAr{0})} \kB ,
 \label{eq:EntropyA}
 \\
 \SM &= - \NumSites^3 \kantpar{\fMr{0} \ln \fMr{0} - (1 + \fMr{0}) \ln(1 + \fMr{0})} \kB .
 \label{eq:EntropyM}
\end{align}
\end{subequations}
We prove in Appendix \ref{AppndS} that at fixed temperature the total entropy $\STot = \SA + \SM$, combining results (\ref{eq:EntropyA}) and (\ref{eq:EntropyM}), is maximized when the resonance energy takes the value $\EresMaxS = 2\EAnum{0} - \EMnum{0}$, and in the deep lattice limit this may be approximated by
\begin{align}
\label{eq:EresSmaxApprox}
\EresMaxS \approx \EF - \af{\frac{3}{8} - \frac{3}{16}\frac{1}{\sqrt{\Pot_{0,\Atom} / \ERA}}} \ERA .,
\end{align}
using (\ref{eq:EnergySplit}) and setting $\EAnum{0} \approx \EF$. This coincides with the resonance energy  at which the system obtains a 50\% molecule fraction at zero temperature \cite{Ole2008a}.

\begin{figure*}[bhtp]
    \centering
    \includegraphics[width=0.8\textwidth]{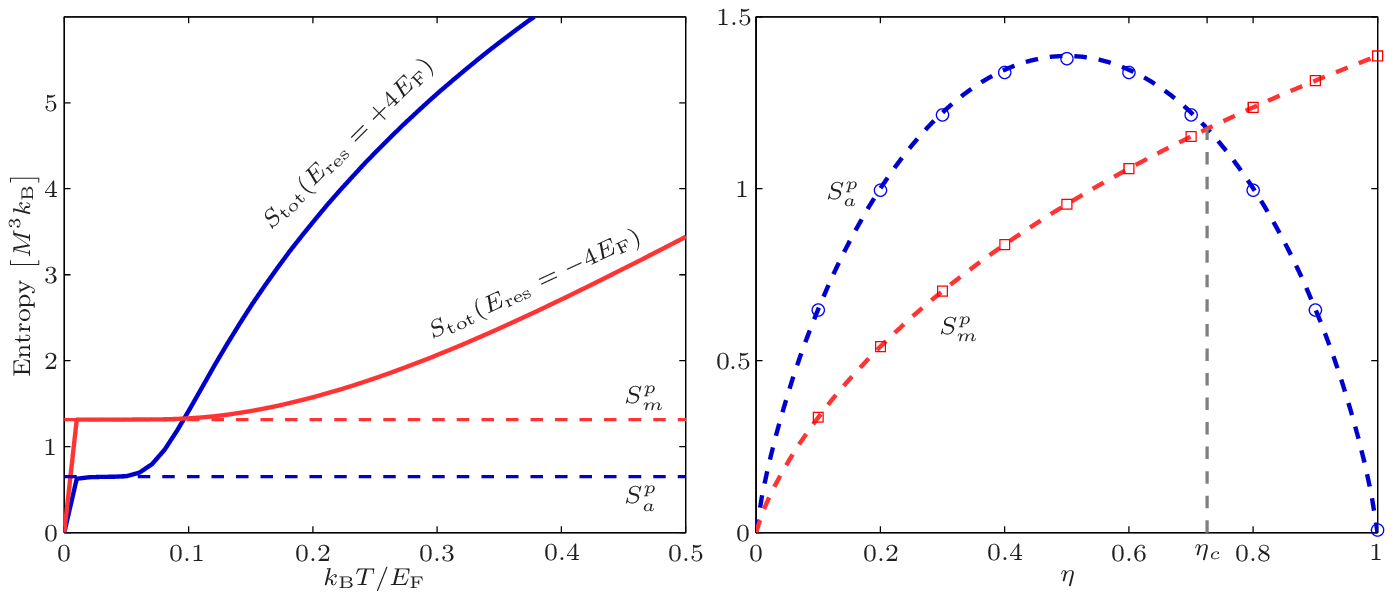}
    \caption{Entropic behavior for $\Pot_0 = 10 \ERA$. Left: The total entropy at $\Eres = - 4 \EF$ (red curve) and $\Eres = + 4 \EF$ (blue curve), corresponding to a pure Bose and Fermi gas, respectively. The curves have plateaus, which are marked by the two dashed horizontal lines. Right: The height of the plateaus are read off manually ({\color{red} $\boldsymbol{\scriptscriptstyle\Box}$} for the molecular and {\color{blue} $\boldsymbol{\circ}$} for the atomic entropy plateau). These are well approximated by Eqs. (\ref{eq:LimitEntropyA}) and (\ref{eq:LimitEntropyM}) (dashed curves) and intersect at the filling fraction $\FFC = 0.726$.}
  \label{fig:EntropiPlateau}
\end{figure*}

In \figurename~\ref{fig:EntropyTop} we present two examples from the numerical analysis of the behavior of the system entropy along an isotherm, and the portion of the system entropy arising from the atomic and molecular subsystems. 
In both cases it is clear that the system entropy is maximized in the transition region at a value of the resonance energy that is in good agreement with our analytic expression for $\EresMaxS$.
The first case of $\FF = 0.5$ shown in \figurename~\ref{fig:EntropyTop} reveals that the maximum in entropy  arises from the rapid growth of degrees of freedom as population is coupled into the new subsystem (i.e. the growth of the atomic (molecular) entropy as we approach the transition from the left (right)) exceeding the reduction of entropy from the subsystem loosing population.

In the second case of $\FF = 0.9$ shown in \figurename~\ref{fig:EntropyTop}, we see behavior not fitting the previous description: not only the molecular, but also the atomic subsystem entropy increases as the transition is approached from the right. 
The key feature leading to this is the non-monotonic dependence of the atomic subsystem entropy on $\Eres$, which occurs when the filling satisfies $\FF > \frac{1}{2}$. We note from (\ref{eq:SingleLevelEntropy}) that $\SA$ vanishes for  $\fAnum{0}=0$ and $\fAnum{0}=1$, and is maximum for $\fAnum{0} = \frac{1}{2}$. Thus for  high filling ($\FF > \frac{1}{2}$) as the resonance energy is lowered and the transition region is approached (from above) $\fAnum{0}$ decreases,  initially leading to an increase in $\SA$, until $\fAnum{0}$  decreases below $\frac{1}{2}$, at which point $\SA$ begins to decrease. 
In contrast, due to their bosonic statistics the molecules give a contribution to the total entropy, which increases with increasing $\fMnum{0}$.

This peak in the atomic entropy can also be understood from (\ref{eq:parddSadfa}), which states that the sign of $\parddt{\SA}{\Eres}$ is determined by the sign of $\EAnum{0} - \KemPot$, and this can be negative if the chemical potential lies higher than the ground atomic energy level. This requires $\frac{1}{2}<\FF\leq1$, since in the deep lattice limit $\KemPot = \EAnum0 - \kB T\ln \left(\frac{1-\FF}{\FF}\right)$ on the atomic side of the resonance, assuming \refEq{eq:Econdition}. Note that there is no analogue effect for the molecules, since we always have $\EM + \Eres < 2\KemPot$ in (\ref{eq:parddSmdfm}) due to Bose-Einstein statistics.

\subsection{Entropy plateaus}\label{sec:EntropyPlateaus}
On the atomic side of the transistion zone ($\Eres\gtrsim2\EF$) and for the range of $T$ where (\ref{eq:Econdition}) is valid, $\fA$ is  constant at the value of $\FF$. The associated entropy is entirely due to the atomic subsystem and is constant at the \emph{plateau} value 
\begin{subequations}
\label{eq:LimitEntropy}
\begin{align}
 \SAP  &= - 2 \NumSites^3 \kantpar{\FF \ln \FF + (1 - \FF) \ln(1 - \FF)} \kB .
 \label{eq:LimitEntropyA}
\end{align}
Also, in a similar temperature regime on the molecular side of the transistion zone  ($\Eres\lesssim-2\EF$) we have a constant molecular occupation $\fM = \FF$ and hence an associated entropy plateau of
\begin{align}
\SMP &= - \NumSites^3 \kantpar{\FF \ln \FF - (1 + \FF) \ln(1 + \FF)} \kB .
\label{eq:LimitEntropyM}
\end{align}
\end{subequations}

These plateaus are indicated in \figurename~\ref{fig:EntropyTop} as $\Eres$ varies and clearly provide a good description of the total system entropy on either side of the transition region.
In \figurename~\ref{fig:EntropiPlateau} we show the total entropy for the system as a function of temperature for $\Eres = + 4 \EF$ (in the atomic regime) and $\Eres = - 4 \EF$ (in the molecular regime), respectively. These results clearly show the importance of the energy scales $\BandWidthThreeD{0}$ and $\GapThreeD{0}$: For $\BandWidthThreeD{0} \ll \kB T \ll \GapThreeD{0}$ the plateau is observed in good agreement with the analytical prediction. For lower temperatures, $\kB T \ll \BandWidthThreeD{0}$, the ground band is not uniformly occupied leading to a sharp suppression of entropy near $T=0$. At higher temperatures, $\GapThreeD{0} \ll \kB T$, excited bands become thermally accessible and contribute additional entropy. 

As is clear from the results in \figurename~\ref{fig:EntropiPlateau} in  the temperature range $\BandWidthThreeD{0}  \ll \kB T \ll \GapThreeD{0}$  the entropy on the plateaus can be quite accurately determined from the full numerical calculations. We compare the results of such values against the analytic expressions (\ref{eq:LimitEntropy})  for a range of filling fractions in \figurename~\ref{fig:EntropiPlateau} on the right. The curves for the atomic and the molecular entropy plateaus intersect at the filling fraction $\FFC$
\begin{align}
\label{eq:FFC}
\FFC = 0.726,
\end{align}
as is seen by equating (\ref{eq:LimitEntropyA}) and (\ref{eq:LimitEntropyM}). 

  \begin{figure}[hb]
    \centering
    \includegraphics[width=\columnwidth]{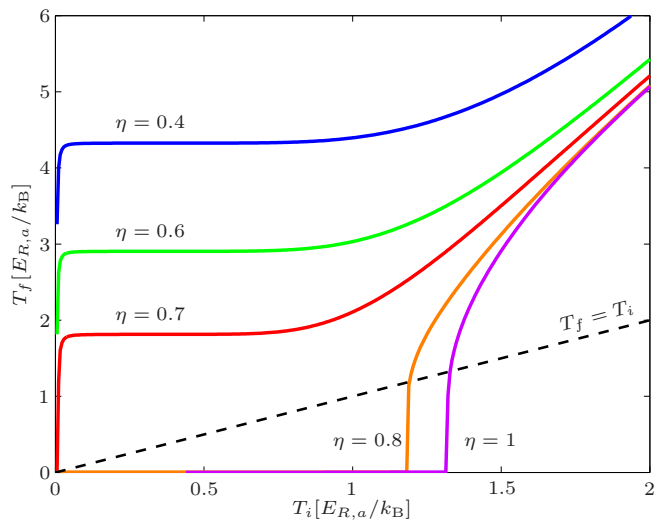}
    \caption{The dependence of the final temperature on the initial temperature in adiabatic $\Eres$-sweeps from the atomic to the molecular regime for different filling fractions and for a lattice depth $\Pot_0 = 20 \ERA$. For $\FF < \FFC$ the temperature increases while it decreases (for sufficiently low $\Ti$) for $\FF > \FFC$.}
  \label{fig:TiTfEresSweep}
\end{figure}

\section{Adiabatic sweep of $\Eres$}
Because ultra-cold atom systems are effectively isolated, i.e. not thermally connected to a reservoir, the various manipulations that can be made to the system will usually result in a change in its equilibrium temperature.  
Such changes in temperature that occur during loading into optical lattices can be appreciable and has been of considerable recent interest \cite{Blakie2004a,Blakie2005a,Kohl2006a,Blakie2007a,Blakie2007b,Ho2007a,Gerbier2007a,Koetsier2008a,Cramer2008a}. 
Here we investigate the temperature changes arising in the Feshbach coupled Fermi gas in an optical lattice. To do this we consider our initial system in the atomic regime with temperature $\Ti$ and entropy $\Si\af{\Ti}$ and then determine the final temperature after an adiabatic sweep of $\Eres$ into the molecular regime by equating the entropy functions, $\Sf(\Tf) = \Si(\Ti)$, where $\Sf$ is the system entropy at the final $\Eres$ value.  
\figurename~\ref{fig:TiTfEresSweep} shows our results for the relationship between $\Ti$ and $\Tf$ for various  filling fractions.
For moderate fillings, $\FF = 0.4,0.6,0.7$, we observe that the sweep leads to a final temperature much greater than the initial temperature. This can be qualitatively understood because there are half as many molecules as atoms, and their temperature needs to be higher to carry the same amount of disorder as the unbound atoms.
Of course this argument is not general. As discussed earlier, in the atomic regime in conditions where only the ground band is accessible, as the filling fraction increases past $\FF=0.5$, the entropy of the fermionic atoms decreases (at fixed $T$) -- on the other hand, the bosonic entropy only increases with increasing $\FF$. Our results show that for the parameters under consideration in \figurename~\ref{fig:TiTfEresSweep} that for $\FF\gtrsim 0.8$, there is a wide region of initial temperatures for which   the Feshbach sweep leads to a much colder molecular gas.

\begin{figure*}[t]
   \centering
   \includegraphics[width=0.8\textwidth]{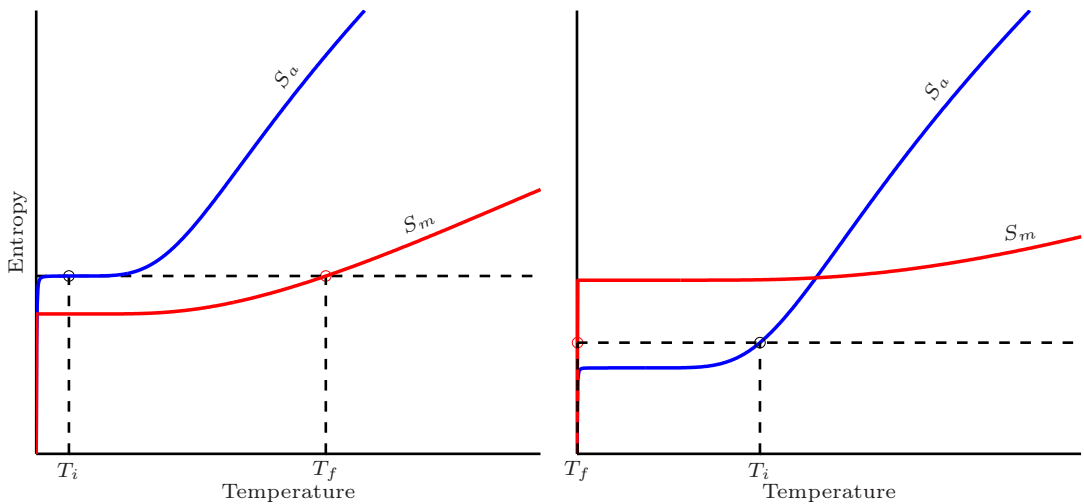}
   \caption{Schematic illustration of the adiabatic cooling. Left: If the atomic entropy plateau lies above the molecular one ($\FF < \FFC$) the final temperature in an adiabatic $\Eres$-sweep will be higher than the initial temperature as shown to the left. However, if the molecular entropy plateau lies above the atomic one ($\FFC < \FF \leq 1$) the final temperature will be comparable to the bandwidth of the molecular spectrum,  $\BandWidthThreeD[\Molecule]{0}$.}
 \label{fig:EntropiPlateauSkematisk}
\end{figure*}

This behavior is quantitatively explained in \figurename~\ref{fig:EntropiPlateauSkematisk} where the entropy curves in the atomic and molecular regimes are shown for two different cases: (1) If the atomic entropy plateau lies above the molecular one ($\FF < \FFC$) and we start with an initial temperature $\Ti$ corresponding to an atomic entropy on the plateau, then $\Tf$ will be independent of and lie above $\Ti$. (2) If on the other hand the atomic entropy plateau lies below the molecular one ($\FFC < \FF < 1$), the final temperature will be much lower, i.e. lie in the regime where $\kB T$ is comparable to $\BandWidthThreeD{0}$. A detailed prediction for the final temperature requires a theory for the molecular entropy dependence on temperature outside of the plateau region. There is no general closed-form expression for this in the lattice potential, however in the regime where significant temperature reduction is predicted the uniform Bose gas expression should be applicable with the replacement of $m_\Molecule$ by an effective mass including the influence of the lattice (it should be noted that in the molecular regime for $\kB T\lesssim \BandWidthThreeD[\Molecule]{0}$ condensation is likely and will need to be included to make a quantitative prediction).

The above results only apply when the the initial state lies on the entropy plateaus (i.e. $\Eres\gtrsim2\EF$ and $\BandWidthThreeD[\Atom]{0} \ll \kB T \ll \GapThreeD[\Atom]{0}$).
The upper  temperature limit, $\GapThreeD[\Atom]{0}$, increases with increasing $\Pot_0$, and the lower limit, $\BandWidthThreeD[\Atom]{0}$, decreases with increasing $\Pot_0$, so that the region of applicability of this simple analysis improves with increasing lattice depth. However, this is also the regime in which interaction effects become more important requiring a description beyond that presented here, if quantative predictions are to be obtained.

\section{The role of interactions}
We have thus far neglected interactions in our analysis to elucidate the aspects of the thermodynamics arising from the band structure in the lattice potential. We now briefly discuss how interactions will modify our conclusions. Our analysis of isentropic sweeps of the resonance energy relied on the existence of plateaus of constant entropy on both sides of the Feshbach resonance. We therefore concentrate on how interactions impact the entropy in the molecular and the atomic limits. 

On the molecular side of the resonance we have a purely bosonic system with weakly repulsive residual interactions. As the lattice depth is increased correlation effects get stronger in the system. For the translationally invariant lattice the special case where the number of bosons is commensurate with the number of lattice sites then the system can undergo a quantum phase transition to the Mott-insulating phase. In the Mott-insulating phase a gap, given by the on-site interaction strength ($U_m$),  emerges in the excitation spectrum.  For temperatures larger than $U_m/\kB$ (but much smaller than the gap to the first excited lattice band) the molecular entropy will still exhibit a plateau~\cite{Rey2006,Koetsier2008,Pollet2008}.  As in the ideal case the existence of an entropy plateau is a consequence of the uniform distribution of the particles over all the states in the lowest band. Hence the value of the molecular entropy on the plateau is determined mostly from combinatorial arguments.     
At lower temperatures the excitation gap exponentially suppresses the entropy of the molecules. 

It is worth noting that the main results of this work have been in the regime $\eta<1$, for which the boson number is incommensurate with the number of lattice sites and the Mott-insulator transition can not occur~\cite{Fisher1989}. For this case the our ideal calculations should have a wider regime of validity.

On the other side of the resonance the interaction between the free atoms is attractive and the ground state is a paired BCS superfluid, characterized by a pairing gap, $\Delta_{\rm{pair}}$. The size of the pairing gap depends on the strength of the attractive interaction $U_a$ compared with the bandwidth $\BandWidthThreeD[\Atom]{}$. In the weak coupling limit $\Delta_{\rm{pair}}$ is exponentially small and proportional to $\BandWidthThreeD[\Atom]{}$, while for stronger interactions the size of $\Delta_{\rm{pair}}$ is set by $|U_a|$~\cite{Nozieres1984}. As was the case for the molecules the atomic entropy is  reduced by an exponential factor for $\kB T$ smaller than $\Delta_{\rm{pair}}$, while it reaches a constant plateau in the intermediate temperature regime $\Delta_{\rm{pair}}<\kB T\ll \GapThreeD[\Atom]{}$.
Hence correlated states introduce a new low energy scale $|U_\sigma|$, which replaces the bandwidth $\BandWidthThreeD{0}$ as the relevant energy scale for the existence of an entropy plateau, and our conclusions remain qualitatively correct when interactions are included, provided $|U_a|\ll\Delta_a$ and  $U_m\ll\Delta_m$ in the deep lattice limit. Whether this inequality can be satisfied will depend on the details of the interactions. 

Finally, we remark that while the Feshbach molecules are produced in the least bound ro-vibrational state they are remarkably robust against collisional deexcitation, which could lead to losses when a molecule shares a lattice site with another molecule or a free atom. This stability arises from Pauli blocking of the the constituent fermionic atoms~\cite{Petrov2004}. Again we note that since our primary interest here is in systems with filling fractions less than unity,  the effect of collisional losses will be smaller.

When an external harmonic trapping potential is present the qualitative similarity of the physics of the ideal and strongly correlated regimes is not clear. For instance, in this case the gas will in general contain several distinct spatial regions exhibiting either superfluid or insulating behavior~\cite{Jaksch1998}. Hence this situation will require future work.
\section{Conclusions}

In this paper we have considered the Feshbach resonance based association of a Fermi gas of atoms confined in a uniform optical lattice potential. 
Our analysis of the statistical mechanics of this system has revealed a rich range of behavior in this system arising from the strikingly different degenerate properties of the atom and molecular degrees of freedom, and the well-isolated ground band that forms in moderately deep lattice potentials. 
We have considered the effect of an adiabatic ramp of the Feshbach resonance from the atomic to molecular regimes, and shown that a colder molecular gas can be produced for a wide parameter regime if the filling fraction lies in the range $ 0.726 < \FF < 1$.

By focussing on the ideal gas case we have been able to clarify the effect of the translationally invariant lattice potential on the atom-molecule equilibrium. Building on this study, future work will be to include interactions and the additional confinement present in experiments due to the inhomogeneous harmonic trap potential.

\acknowledgments{N. N. acknowledges financial support by the Danish Natural Science Research Council.} 

\appendix 

\section{Analytic expressions for the first band gap}\label{AppndE}

In the deep lattice limit we can make a tight-binding approximation and regard the potential as a collection of $\NumSites^3$ independent wells. The central band energy is then given by the discrete energy levels $\Einum{n}$ in each well which to lowest order are harmonic oscillators, but anharmonic corrections can be calculated using perturbation theory (see Appendix A of Ref. \cite{Ole2008a}). Coupling between wells induces a finite bandwidth, that we neglect in the deep lattice (flat band) limit.
Because the lattice potential is separable the energies are of the form 
\begin{align}
\Einum{n}=\EOneD_\Type^{n_x}+ \EOneD_\Type^{n_y}+\EOneD_\Type^{n_z},
\label{eq:3DEnergyLevels}
\end{align}
where we refer to the $\EOneD_\Type^{n_j}$ as the 1D energies along the $x_j$ direction, and $\{n\}\leftrightarrow\{n_x,n_y,n_z\}$ are the quantum numbers describing the single well state. The lowest two 1D energy levels are given by~\cite{Ole2008a}
\begin{subequations}
\label{eq:En1D}
\begin{align}
 \label{eq:Ezero1D}
 \EOneD^0_\Type &\approx \af{\sqrt{\frac{\Pot_{0,\Type}}{\ERi}} -
   \frac{1}{4} - \frac{3}{48} \frac{1}{\sqrt{\Pot_{0,\Type} /
       \ERi}}} \ERi,\\
 \label{eq:Eone1D}
 \EOneD^1_\Type &\approx \af{3\sqrt{\frac{\Pot_{0,\Type}}{\ERi}} -
   \frac{5}{4} - \frac{49}{48} \frac{1}{\sqrt{\Pot_{0,\Type} /
       \ERi}}} \ERi .
\end{align}
\end{subequations}
From (\ref{eq:3DEnergyLevels}) we see that the lowest 3D energy band lies at $\Einum{0} = 3\EOneD_\Type^0$ while the next 3D energy band lies at $\Einum{1} = 2\EOneD_\Type^0 + \EOneD_\Type^1$ yielding a bandgap of
\begin{align}
\GapThreeD{0}= \Einum{1} - \Einum{0} = \EOneD_\Type^1 -
\EOneD_\Type^0 .
\end{align}
Approximating the two lowest 1D energy levels by (\ref{eq:En1D}) we have
\begin{align}
\label{eq:GapAppr}
\GapThreeD{0}
&\approx \af{2\sqrt{\frac{\Pot_{0,\Type}}{\ERi}} - 1 - \frac{23}{24} \frac{1}{\sqrt{\Pot_{0,\Type} / \ERi}}} \ERi .
\end{align}

\section{Maximising entropy in the flat-band limit}\label{AppndS}
We are interested in how the total entropy $\STot = \SA + \SM$ changes if we change the resonance energy but keep the temperature constant, i.e.
\begin{align}
\label{eq:dStotdEres}
\ddconst{\STot}{\Eres}{T}
&= \pardd{\SA}{{\fAr{0}}} \ddconst{{\fAr{0}}}{\Eres}{T} + \pardd{\SM}{{\fMr{0}}} \ddconst{{\fMr{0}}}{\Eres}{T} .
\end{align}
The number of unbound atoms is $\NumA = 2 \NumSites^3 \fAr{0}$ and the number of molecules is $\NumM = \NumSites^3\fMr{0}$, so by combining (\ref{eq:FFdef}) with the condition of particle conservation (\ref{eq:NumCons}) we get
\begin{align}
\label{eq:fAfMFF}
\fAr{0} + \fMr{0} = \FF .
\end{align}
We may thus rewrite (\ref{eq:dStotdEres}) as
\begin{align}
\label{eq:dStotdEres2}
\ddconst{\STot}{\Eres}{T}
&= \tuborg{\pardd{\SA}{{\fAr{0}}} - \pardd{\SM}{{\fMr{0}}}} \ddconst{\fAr{0}}{\Eres}{T}.
\end{align}
From (\ref{eq:SingleLevelEntropy}) the partial derivatives of $\SA$ and $\SM$ with respect to the occupation numbers $\fAr{0}$ and $\fMr{0}$ can be found, leading to
\begin{subequations}
\begin{align}
 \pardd{\SA}{{\fAr{0}}}
 &= \frac{2 \NumSites^3(\EAnum{0} - \KemPot)}{T} ,
 \label{eq:parddSadfa}
 \\
 \pardd{\SM}{{\fMr{0}}}
 &= \frac{\NumSites^3(\EMnum{0} + \Eres - 2\KemPot)}{T} ,
 \label{eq:parddSmdfm}
\end{align}
\end{subequations}
which we can insert in (\ref{eq:dStotdEres2}), yielding
\begin{align}
\ddconst{\STot}{\Eres}{T}
&= \frac{\NumSites^3\af{2 \EAnum{0} - \EMnum{0} - \Eres}}{T} \ddconst{\fAr{0}}{\Eres}{T} .
\label{eq:dStotdEresFinal}
\end{align}%

The dependence of $\Eres$ on $\fAr{0}$ is via the chemical potential, so the derivative  is
\begin{align}
\ddconst{\fAr{0}}{\Eres}{T}
&= \frac{1}{\kB T} \frac{\Exp{(\EAnum{0} - \KemPot)/\kB T}}{(\Exp{(\EAnum{0} - \KemPot)/\kB T} + 1)^2} \pardd{\KemPot}{\Eres} ,
\end{align}
which is positive since $\parddt{\KemPot}{\Eres} > 0$ in this regime~\cite{Ole2008a}. Thus we see from (\ref{eq:dStotdEresFinal}) that the sign of the derivative of $\STot$ with respect to $\Eres$ is the same as the sign of the quantity $2\EAnum{0} - \EMnum{0} - \Eres$. We conclude that the entropy is maximal when the resonance energy equals
\begin{align}
\EresMaxS \defi 2\EAnum{0} - \EMnum{0} = \EAnum{0} - (\EMnum{0} - \EAnum{0})
\end{align}
and in this deep lattice limit we can approximate $\EMnum{0} - \EAnum{0}$ by (\ref{eq:EnergySplit}) and set $\EAnum{0}$ equal to the Fermi energy, such that we get
\begin{align}
\EresMaxS \approx \EF - \af{\frac{3}{8} - \frac{3}{16}\frac{1}{\sqrt{\Pot_{0,\Atom} / \ERA}}} \ERA .
\end{align}

\end{document}